\documentclass[12pt]{article}
\usepackage{bm}
\usepackage[fleqn]{amsmath}
\usepackage{hyperref}
\usepackage{graphicx}
\usepackage{lipsum}
\usepackage{amssymb}
\usepackage{multirow}

\begin{document}

\title{Piezoelectric networks and ferroelectric domains in twistronic superlattices in WS$_2$/MoS$_2$ and WSe$_2$/MoSe$_2$ bilayers}

\author
{
V.\,V. Enaldiev$^{1,2,3*}$ F. Ferreira$^{1,2}$, S.\,J. Magorrian$^{1,2}$\\ 
Vladimir I. Fal'ko,$^{1,2,4*}$
\\
\\
\normalsize{$^{1}$School of Physics and Astronomy, University of Manchester, Oxford Road,}\\
\normalsize{ Manchester, M13 9PL, UK}\\
\normalsize{$^{2}$National Graphene Institute, University of Manchester, Oxford Road,}\\
\normalsize{ Manchester, M13 9PL, UK}\\
\normalsize{$^{3}$Kotel'nikov Institute of Radio-engineering and Electronics,} \\ 
\normalsize{Russian Academy of Sciences, 11-7 Mokhovaya St, Moscow, 125009 Russia}\\
\normalsize{$^{4}$Henry Royce Institute for Advanced Materials, University of Manchester,}\\
\normalsize{Oxford Road, Manchester, M13 9PL, UK}\\
\\
\normalsize{ E-mail: vladimir.enaldiev@manchester.ac.uk, vladimir.falko@manchester.ac.uk}
}

\date{\today}

\maketitle

{\bf 
Twistronic van der Waals heterostrutures offer exciting opportunities for engineering optoelectronic properties of nanomaterials, in particular, due to the formation of moir\'e superlattice structures. In twisted bilayers of transition metal dichalcogenides moir\'e superlattice effects are additionally enriched by the lack of inversion symmetry in each monolayer unit cell. Here, we use multiscale modeling to establish a rich variety of confinement conditions for electrons, holes and layer-indirect excitons in twistronic WX$_2$/MoX$_2$ bilayers (X=S,Se). Such trapping of charge carriers and excitons is caused by ferroelectric (interlayer) polarisation and piezoelectric effects generated by the reconstruction of twistronic bilayers into preferential stacking domains separated by domain wall networks. For almost aligned bilayers with anti-parallel (AP) orientation of WX$_2$ and MoX$_2$ unit cells, we find that upon lattice relaxation piezoelectric potential modulation traps holes and electrons in the opposite corners -- WMo and XX (tungsten over molybdenum {\it versus} overlaying chalcogens) -- of hexagonal-shaped 2H (simultaneously WX and XMo) stacking domains, swapping their positions at a twist angle $\gtrsim 0.2^{\circ}$. This crossover happens at such small angles (set by a very small lattice mismatch between WX$_2$ and MoX$_2$) that would impose an alignment accuracy and homogeneity better than $0.1^{\circ}$ for achieving reproducibility of electronic characteristics of such heterostructures. At the same time, for all angles, XX corners provide $30$\,meV deep traps for the interlayer excitons. In bilayers with parallel (P) orientation of WX$_2$ and MoX$_2$ unit cells, band edges for both electrons and holes appear in triangular domains, where WX$_2$ chalcogens set over MoX$_2$ molybdenums. We find that, due to a weak ferroelectric polarisation, these triangular domains act as $130$\,meV deep quantum boxes for interlayer excitons for twist angles $\lesssim 1^{\circ}$, shifting towards XX stacking sites of the domain wall network at larger twist angles.     
}

\begin{figure}
	\includegraphics[width=1.\columnwidth]{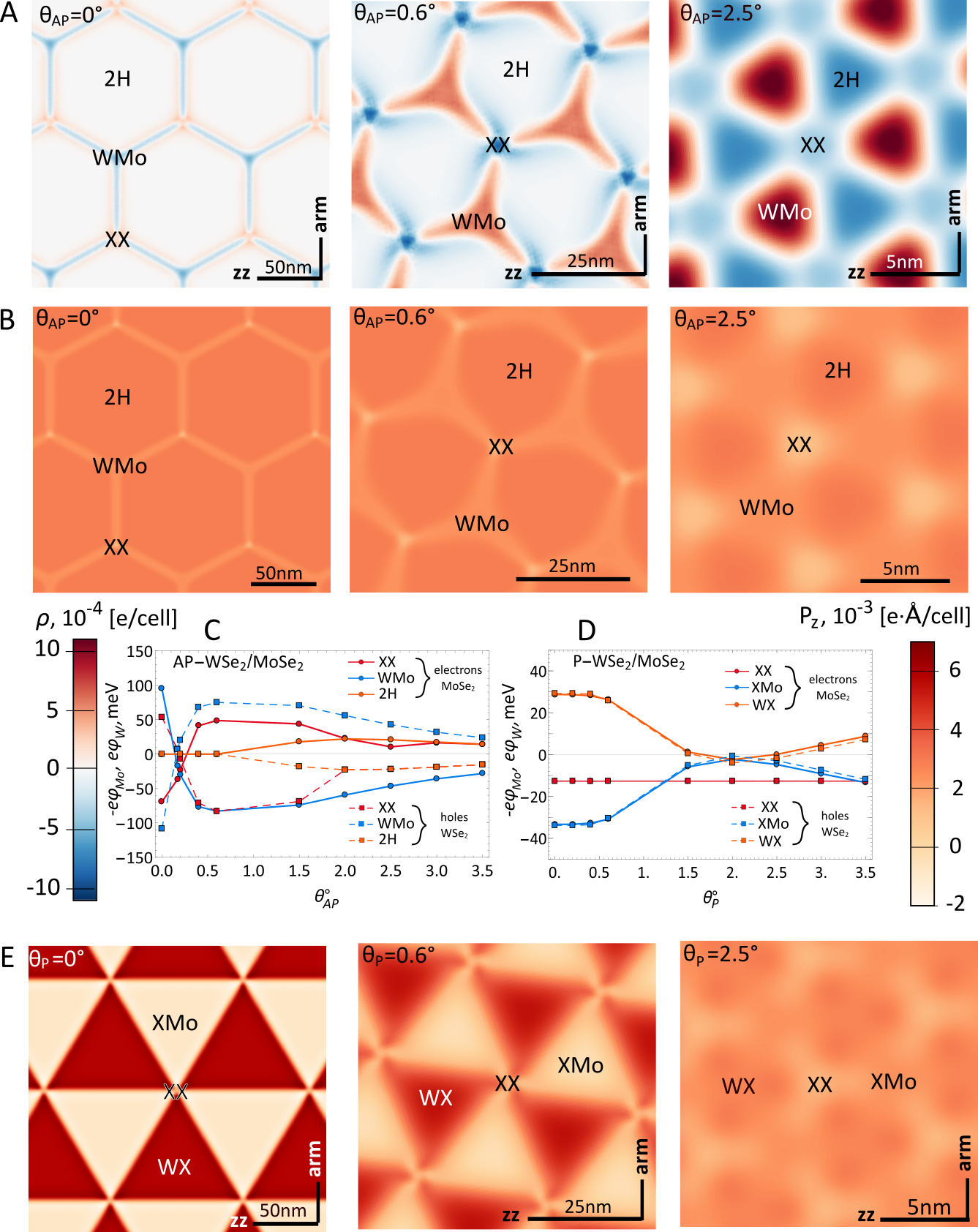}
	{\small\caption{\label{Fig0} {\bf Intra- and interlayer charge transfers in twisted WSe$_2$/MoSe$_2$ bilayers.} {\bf A}  Maps of piezocharge $\rho$ and {\bf B} out-of-plane electric polarisation $P_z$ densities in AP-WSe$_2$/MoSe$_2$ bilayers with  labeled twist angles. {\bf C,D.} Twist angle dependences of band edge energies of electrons and holes {\bf C} in XX, WMo, 2H  regions for AP- and {\bf D} in XX, XMo, WX regions for P-WSe$_2$/MoSe$_2$ bilayers. {\bf E.} Maps of out-of-plane polarisation, $P_z$, for P-WSe$_2$/MoSe$_2$ bilayers for the same twist angles as for AP-bilayers. Panels on {\bf B} and {\bf E} have the same color bar scale. WS$_2$/MoS$_2$ bilayers show a qualitatively similar behaviour. For all maps horizontal and vertical directions correspond to zigzag and armchair crystallographic axes of constituent monolayers, respectively.  
	}}
\end{figure}

Moir\'e superlattices are generic for van der Waals structures assembled from two-dimensional materials (2DM): graphene \cite{cao2018unconventional,cao2018correlated}, hexagonal boron nitride \cite{Ponomarenko2013,dean2013hofstadter,LeeScience2016}, and monolayers of transition metal dichalcogenides (TMDs) \cite{kunstmann2018momentum,rivera2018interlayer,nayak2017probing,alexeev2019resonantly,tran2019evidence}. For pairs of 2DM with a small lattice mismatch and ``twistronic'' bilayers with a small misalignment angle, $\theta$, between the crystallographic axes moir\'e superlattices have long periods, $\ell$, giving rise to minibands for charge carriers \cite{cao2018unconventional,cao2018correlated,Ponomarenko2013} and excitons \cite{kunstmann2018momentum,rivera2018interlayer,nayak2017probing,alexeev2019resonantly,tran2019evidence}.  For twistronic structures with the longest periods, local lattice reconstruction occurs into an array of the preferential stacking domains corresponding to the equilibrium structure of bulk van der Waals crystals \cite{rosenberger2020twist,Weston2020,alden2013strain,yoo2019atomic}. While lattice reconstruction is stronger in marginally twisted homobilayers of 2D materials (graphene on graphene, or MX$_2$/MX$_2$ \cite{yoo2019atomic,Weston2020,Edelberg2020}), the very close lattice constants of TMDs with different metals but the same chalcogen (WX$_2$ and MoX$_2$) also promote the formation of preferential stacking domains in marginally twisted heterobilayers \cite{Weston2020,rosenberger2020twist}. Upon lattice reconstruction, 
moir\'e pattern in such structures takes the form of large-area ($\propto \ell^2$) domains separated by a network of dislocation-like domain walls \cite{Weston2020,Edelberg2020,rosenberger2020twist,Enaldiev_PRL,PRLNaik,carr2018relaxation}. Due to the inversion asymmetry of TMD monolayers, the emerging domain structures differ for bilayers with parallel (P, $\theta\equiv\theta_P$) and anti-parallel (AP, $\theta=\pi + \theta_{AP}$) orientations of their unit cells. While for P-bilayers the reconstructed moir\'e pattern constitutes triangular domains of 3R-type stacking (either WX, or XMo), Fig. \ref{Fig0}, domains in AP-bilayers are hexagonal and feature 2H-type stacking (simultaneously, WX and XMo) \cite{Weston2020,rosenberger2020twist,Enaldiev_PRL,PRLNaik}. 
  
Here, we demonstrate that such TMD heterostructures exhibit a rich variety of quantum dot-like features for electrons ($e$), holes ($h$) and interlayer excitons ({{\it iX}}) which positions within moir\'e superlattices strongly depend on the misalignment angle. Because MoX$_2$/WX$_2$ heterostructures have a type-II band alignment, with the conduction band edge residing in MoX$_2$ and the valence band edge in WX$_2$, the interlayer hybridization of the band edges is non-resonant and is superseded by electrostatic potentials due intralayer piezoelectric and interlayer ferroelectric charge transfers. The resulting variety of features is illustrated in Fig. \ref{Fig0} for  WSe$_2$/MoSe$_2$ heterostructures with both P- and AP-orientations of the unit cells and various small twist angles. For AP-heterostructures, strain concentrated at the domain wall network generates piezocharges with the same signs in the two layers \cite{Enaldiev_PRL}, forming distributions shown in Fig. \ref{Fig0}A. This leads to an electrostatic modulation of the electron (in MoX$_2$) and hole (in WX$_2$) band edges ($\varphi^e_{\rm Mo}$ and $\varphi^h_{\rm W}$), with quantum dots forming in XX corners for electrons and in WMo corners for holes for $|\theta_{AP}^{\circ}|<0.1^{\circ}$. This arrangement swaps to electron/hole band edges at WMo/XX domain wall vortices for $0.3^{\circ}<|\theta_{AP}^{\circ}|\lesssim 2^{\circ}$. To compare, lattice reconstruction in heterostructures with the same orientation of monolayer unit cells (P-bilayers) produces piezoelectric charges with opposite signs in WX$_2$ and MoX$_2$ monolayers \cite{Enaldiev_PRL}, leading to an interlayer polarisation of domain walls, which interplays with the dominant effect -- the interlayer ferroelectric charge transfer oscillating across moir\'e supercell, Fig. \ref{Fig0}E. For twist angles $|\theta_P^{\circ}|\lesssim 1^{\circ}$, these XMo domains act as quantum boxes for electrons, holes and interlayer excitons, Fig. \ref{Fig0}D. For $|\theta_P^{\circ}|\gtrsim 1^{\circ}$, the piezoelectric interlayer potential (extended into smaller domains from domain walls) shifts the electron/hole band edges towards XX stacking areas, followed by a gradual decrease of the band edge modulation across the moir\'e supercell and to the {\it iX} delocalisation at larger twist angles.

\begin{figure}
	\includegraphics[width=1.0\columnwidth]{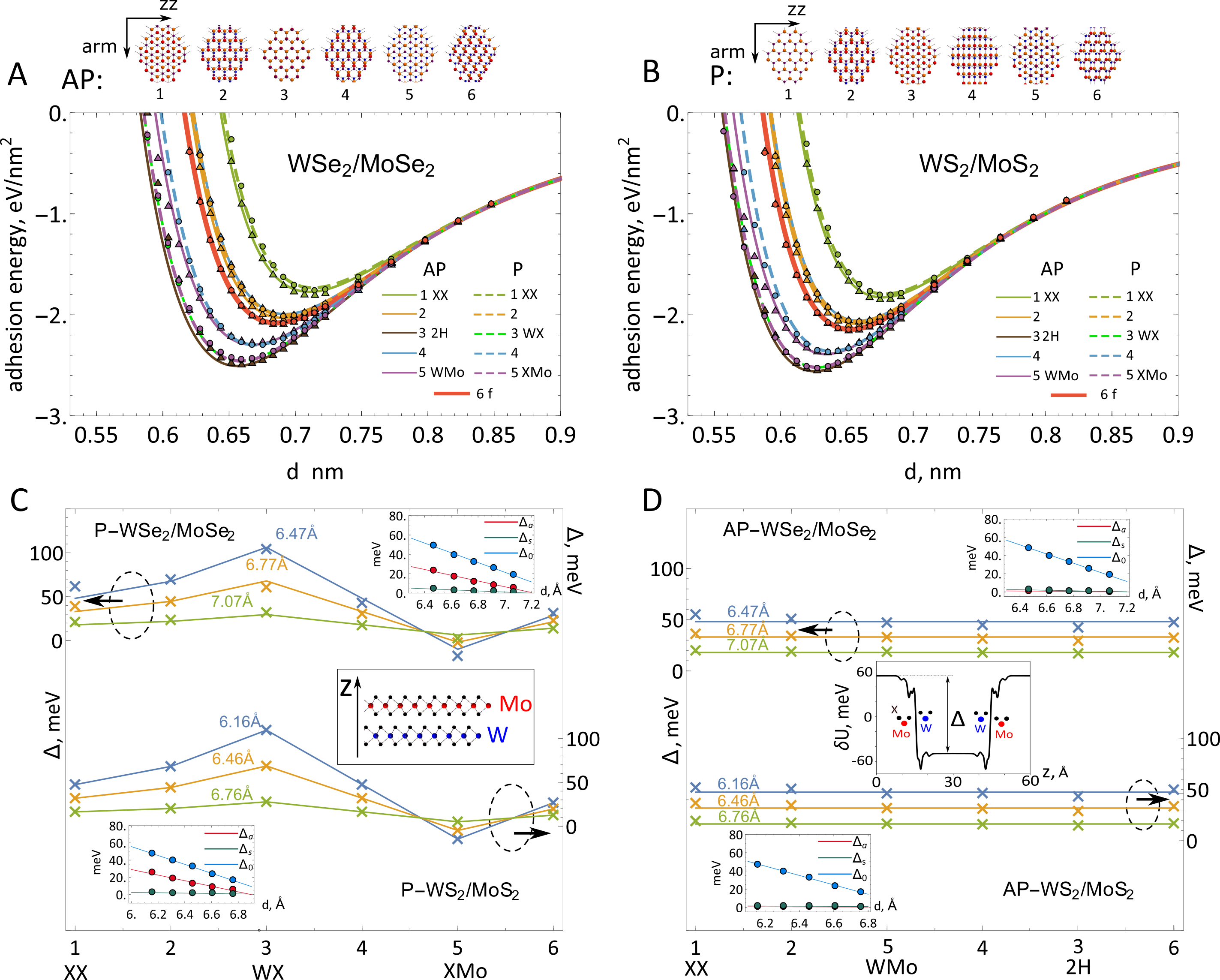}
	{\small \caption{\label{Fig2} {\bf DFT results for adhesion energy, ferroelectric polarisation, and their analysis.} {\bf A} and {\bf B}. Triangles/circles show DFT-computed $d$-dependences of adhesion energies of AP/P-bilayers for six stacking configurations of WX$_2$/MoX$_2$ shown in top insets; solid and dashed lines represent fit for AP and P bilayers; {\bf C.} DFT-computed values (crosses) and their fit (solid lines) with equation (\ref{Eq:potential_difference}) for electron potential energy shift between MoX$_2$ and WX$_2$ monolayers in P-WX$_2$/MoX$_2$ bilayers for three representative interlayer distances. Insets show the corresponding $d$-dependence of $\Delta_{0,a,s}$ and their linear fit. {\bf D.} The same as in {\bf C}, but for AP-WX$_2$/MoX$_2$ bilayers. Middle inset: DFT-computed difference between unit cell-averaged potential energy of a WSe$_2$/MoSe$_2$ heterobilayer (with 3R stacking and distance $d=6.477$\AA), and that of isolated monolayers. The values of $\Delta_{i}$ in {\bf C} and {\bf D} shows that $\Delta_{0}^{AP}=\Delta_{0}^P$ with a high accuracy and $\Delta_s\ll\Delta_0$.
	}}
\end{figure}

The theory proposed in this paper is based on the multiscale analysis \cite{jung2015origin,Enaldiev_PRL} of atomic reconstruction of twisted bilayers using elasticity theory combined with density functional theory (DFT) modeling of interlayer adhesion energy and ferroelectric charge transfer (see Methods). In Fig. \ref{Fig2}A,B, we gather the earlier published DFT results on adhesion energy density \cite{Enaldiev_PRL}, $W(\bm{r}_0,d)$, for WSe$_2$/MoSe$_2$ and WS$_2$/MoS$_2$ heterostructures. Those are presented for both P- and AP-orientations (circles and triangles, respectively) and various stacking configurations characterised by lateral offsets, $\bm{r}_0$, of WX$_2$ and MoX$_2$ honeycomb lattices ($\bm{r}_0=0$ for XX stacking) and interlayer distances, $d$. These DFT data are also encoded into interpolation formulae for $W(\bm{r}_0,d)$, as described in Methods, following the recipes of Refs. \cite{Enaldiev_PRL} and \cite{Weston2020}. For P-oriented structures the most energetically favourable are local stackings with tungstens in WX$_2$ overlaying chalcogens in MoX$_2$ (WX), and molybdenums in MoX$_2$ overlaying chalcogens in WX$_2$ (XMo). Using analogy with the classification of aligned phases of bulk TMDs we will refer to WX and XMo domains as 3R-type domains. For AP-orientation the most favourable is stacking with tungstens in WX$_2$ vertically aligned with chalcogens in MoX$_2$ and molybdenums in MoX$_2$ with chalcogens in WX$_2$ (similar to 2H-stacking in bulk TMDs). Their energies should be compared to the configuration-averaged adhesion energy, $f(d)$, which has a minimum at the interlayer distance $d_0$ (configuration 6). The adhesion energy gained due to forming 2H-type and 3R-type domains, as compared to the minimum $f(d_0)$, is about $\approx 0.5$\,eV/nm$^2$, which is enough to overcome the costs of the uniform strain, $\left(\lambda_{\rm W}+\lambda_{\rm Mo}\right)\delta^2\approx 10$\,meV/nm$^2$, needed for adjusting the lattice constants of perfectly aligned WX$_2$ and MoX$_2$ monolayers ($\delta=0.2\%$ for WS$_2$/MoS$_2$ and $\delta=0.3\%$ for WSe$_2$/MoSe$_2$). This is why moir\'e superlattice of slightly twisted WX$_2$/MoX$_2$ bilayers transforms into 2H-type domains for AP stacking, and into 3R-type domains for P stacking. These domains persist up to angles $\theta_{P}^{\circ}\sim 2.5^{\circ}$ and $\theta_{AP}^{\circ}\sim 1^{\circ}$, whereas for larger twist angles their areas become too small to compensate for the energy cost of the dislocation-like domain wall network.   

The interlayer hybridisation of valence band states in one layer with conduction band states in the other (across the entire band structure) gives rise to the interlayer charge transfer, which can be considered as a weak ferroelectric effect. Such a possibility was discussed in relation to homobilayers of TMDs \cite{Li2017} and was recently observed and analysed in twisted hBN flakes \cite{stern2020,woods2020,yasuda2020}. The formation of a charge double layer at MoX$_2$/WX$_2$ interface produces potential energy jump $\Delta$, quantified in Figs. \ref{Fig2}C and \ref{Fig2}D for P- and AP-bilayers, respectively.  To analyze such a charge transfer by means of DFT modeling we constructed a supercell containing two mirror-reflected WX$_2$/MoX$_2$ heterobilayers placed 30~\AA~apart (to scale down interaction between them) and computed $\Delta$ for P- and AP-WX$_2$/MoX$_2$ heterostructures in the same stacking configurations (same $\bm{r}_0$) as in Fig. \ref{Fig2}A,B across the range of interlayer distances that covers the minima of $W(\bm{r}_0,d)$ for these configurations. All these DFT data are accurately described by an interpolation formula,
\begin{equation}\label{Eq:potential_difference}
	\Delta(\bm{r}_0,d) = A_{0}\left[1-q_0z\right] + A_{a}^{P/AP}\left[1-q_a z\right]\sum_{n=1,2,3}\sin\left(\bm{G}_n\bm{r}_0\right).
\end{equation} 
Here, $z=d-d_0$ is a deviation of the interlayer distance from the minimum of $f$ (configuration average of $W$). The first term in equation (\ref{Eq:potential_difference}) is specific for heterobilayers ($A_0=0$ for homobilayers), and its influence on the indirect (interlayer) band gap comes from the variation of the interlayer distance across the moir\'e pattern. The second term is generic for both homo- and heterobilayers in P-orientation and broken $z\to-z$ symmetry ($\bm{r}_0\neq 0$), but $A_{a}^{AP}=0$ (for homobilayers, this is due to the exact inversion symmetry of their lattice). The values of parameters in this interpolation formula are listed in Table \ref{tab_Fit}. This interpolation formula emerged from fitting the DFT data using a more general periodic function, $\Delta=\Delta_0+\Delta_{a}\sum\sin(\bm{G}_n\bm{r}_0)+\Delta_{s}\sum\cos(\bm{G}_n\bm{r}_0)$, involving the first star ($\pm\bm{G}_{1,2,3}$) of reciprocal lattice vectors of the TMD monolayer, $\Delta_{i}=A_{i}e^{-q_{i}(d-d_0)}\approx A_{i}\left[1-q_{i}(d-d_0)\right]$. We find that $\Delta_s,\Delta^{AP}_a\ll\Delta_0$ (see insets in Fig. \ref{Fig2}C,D). Furthermore, the DFT analysis of the atomic-scale potential variations, shown in the inset in Fig. \ref{Fig2}D, indicates that ferroelectric charge transfer happens between the two inner chalcogen sublayers, which we use to compute interlayer polarisation maps in Fig. \ref{Fig0}.   

{\setlength{\tabcolsep}{2.5pt}
\begin{table}[!h]
\begin{center}
{\small	\caption{Parameters used in the interpolation formulae for ferroelectric charge transfer (equation (\ref{Eq:potential_difference})); for adhesion energy density parameters of WX$_2$/MoX$_2$ bilayers corresponding to DFT data in Fig. \ref{Fig2} (see Methods); elastic moduli \cite{iguiniz2019,androulidakis2018} of the WX$_2$ and MoX$_2$ monolayers used in lattice relaxation and their piezocoefficients \cite{zhu2015observation,rostami2018piezoelectricity}.\label{tab_Fit}}}
	{\small
	\begin{tabular}{c|cccc}
		\hline
		\hline
		X  & $A_0$, & $q_0$,  &  $A_a$,  & $q_a$  \\
		&  meV &  \AA$^{-1}$ &   meV  &  \AA$^{-1}$ \\
		\hline 
		S & 29.9 & 1.73  &  12.8  & 2.53 \\
		Se & 26.5 & 1.89 & 9.53  & 3.12 \\
		\hline
		\hline
        X& \mbox{$w_{1}$}, & \mbox{$w_{2}$}, & $Q$, & $d_0$, \\
        &  eV/nm$^{2}$ &  eV/nm$^2$ & \AA$^{-1}$ &  \AA \\
        \hline
        S &  0.175 & 0.021 &  3.07 & 6.5 \\
        Se & 0.128 & 0.02 &  2.93 & 6.9 \\
        \hline
        \hline
       X & $\varepsilon$,   & $\lambda_{\rm W}$/$\lambda_{\rm Mo}$, & $\mu_{\rm W}$/$\mu_{\rm Mo}$, & $\left|e_{11}^{\rm W}\right|/\left|e_{11}^{\rm Mo}\right|$, \\
        & eV/nm$^4$  &  N/m & N/m & $10^{-10}$C/m \\
        \hline
        S & 214  & 52.5/83.2 & 72.5/70.9 & 2.74/2.9\\
        Se & 189  & 29.7/42.3 & 48.4/49.6 & 2.03/2.14 \\
		\hline
		\hline
	\end{tabular}
	}
	\end{center}
\end{table}
}

\begin{figure}
	\includegraphics[width=\columnwidth]{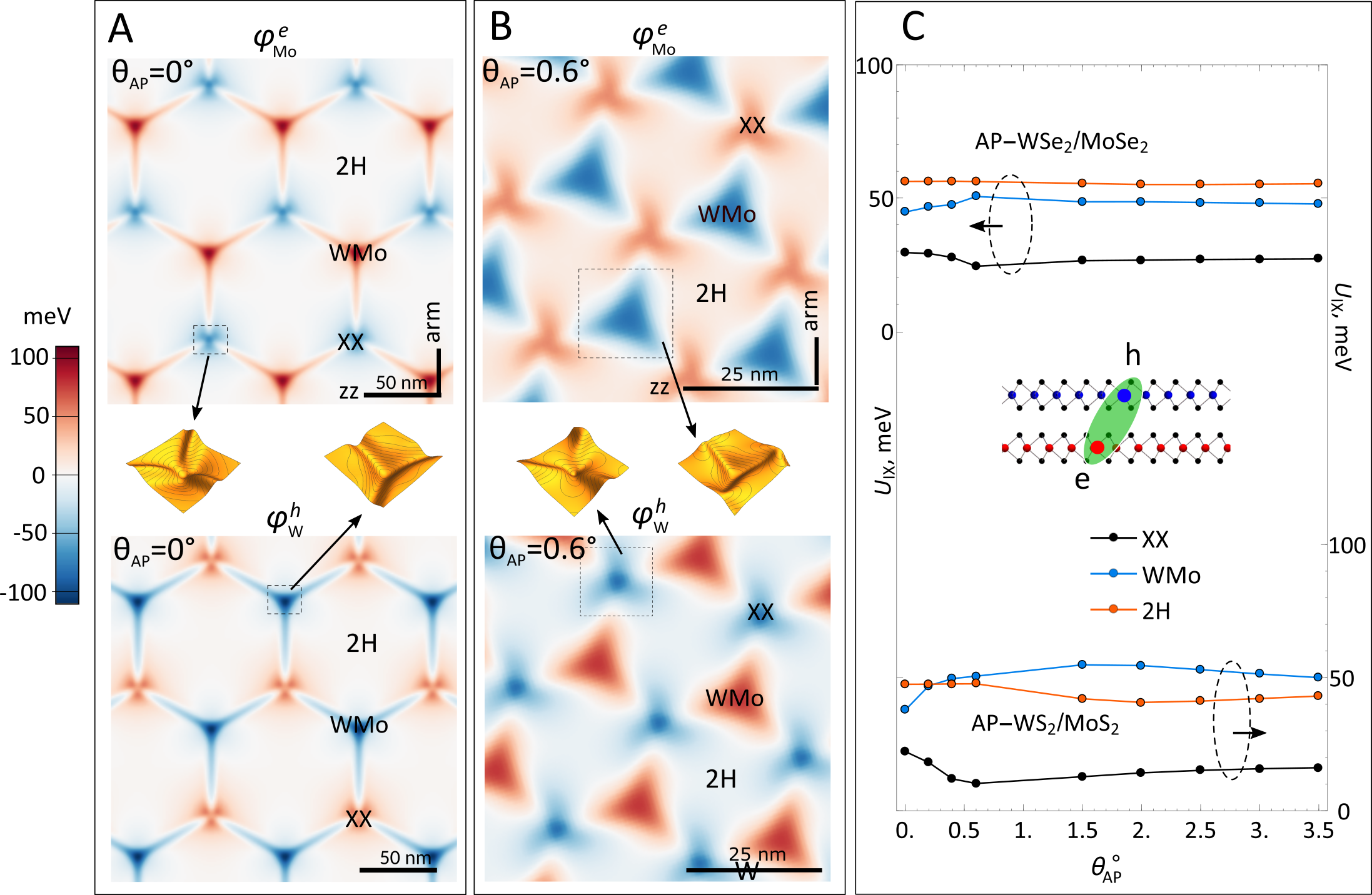}
	{\small \caption{\label{Fig4} {\bf Energy profiles for electrons, holes and interlayer excitons in AP-bilayers.} {\bf A,B.} Maps of electron (top) and hole (bottom) band edge energies for twisted AP-WSe$_2$/MoSe$_2$ bilayers with labeled twist angles. Insets illustrate shape of quantum dot potentials at marginal twist angles. Horizontal and vertical directions correspond to zigzag and armchair crystallographic directions of constituent monolayers, respectively. {\bf C.} Twist-angle-dependence of interlayer exciton energies in XX, WMo, and 2H regions of AP-WSe$_2$/MoSe$_2$ and AP-WS$_2$/MoS$_2$ bilayers. 
	}}
\end{figure}

Lacking inversion symmetry, MoX$_2$ and WX$_2$ monolayers are piezoelectric crystals, so that inhomogeneous strain in each of them produces piezocharges with density, $\rho=e_{11}\left[2\partial_xu_{xy}+\partial_y(u_{xx}-u_{yy})\right]$. It is important to note that piezoelectric charge transfer leads to the most pronounced effects specifically in bilayers with AP-orientation of the unit cells. This is because atomic reconstruction towards 2H-type stacking domains brings together the lattices of the two layers, so that their strain tensors are $u^{\rm W}_{ij}\approx -u_{ij}^{\rm Mo}$, whereas for AP-bilayers the signs of WX$_2$ and MoX$_2$ piezocoefficients are inverted, $e_{11}^{\rm W}\approx-e_{11}^{\rm Mo}$. As a result, piezocharges in the two layers have the same sign and add up, creating a potential varying across moir\'e supercell simultaneously on the top and bottom layers. The cumulative effect of strain and ferroelectric charge transfer on the electron and hole band edges in the AP-bilayers is illustrated in Fig. \ref{Fig4} using WSe$_2$/MoSe$_2$ heterostructures, as an example. The results shown in Fig. \ref{Fig4}A were obtained neglecting interlayer hybridisation of (conduction-conduction and valence-valence) band edge states, which are non-resonant for WX$_2$/MoX$_2$ bilayers (in contrast to homobilayers and some MX$_2$/M$'$X$'_2$ heterostructures with almost resonant band edges, where it plays an important role \cite{alexeev2019resonantly,ruiz2019interlayer}). For very small twist angles, $|\theta_{AP}^{\circ}|<0.1^{\circ}$, electron/hole band edges appear at XX/WMo vortices of the domain wall network, which creates strongly triangulated quantum dot confinement for the corresponding charge carriers, see insets in Fig. \ref{Fig4}A. The electron/hole states localised on a periodic array of quantum dots would produce a narrow band (creating conditions for strongly correlated phases in n- and p-doped bilayers, like those observed in graphene superlattices \cite{cao2018unconventional,cao2018correlated}). For $0.3^{\circ}<\theta_{AP}^{\circ}\lesssim2^{\circ}$, the conduction and valence band edges swap position, with quantum dots for electrons/holes appearing in WMo/XX stacking areas, with the computed confinement profiles shown in Fig. \ref{Fig4}B insets. Emergence of deep trapping potentials for charge carriers in AP-WSe$_2$/MoSe$_2$ bilayers induced by lattice reconstruction has been recently demonstrated in Ref. \cite{shabani2020}. The quantitatively smaller band edge variations caused by the ferroelectric effect additionally modulate the layer-indirect band gap in these type-II heterostructures. This is taken into account in the maps shown in Fig. \ref{Fig4}, though it does not change the positions of the conduction band minima and valence band maxima. For larger misalignment angles $\theta_{AP}^{\circ} \gtrsim 2^{\circ}$, where lattice reconstruction is weak \cite{Enaldiev_PRL,Weston2020}, the piezo- and ferroelectric contributions become comparable, producing a harmonic variation of electron's (in MoX$_2$) and hole's (in WX$_2$) band edges, 
\begin{align}\label{Eq:electron_band_edge_AP}
    \varphi^e_{\rm Mo} = -U_0\sum_{n=1,2,3}\sin\left(\bm{g}_n\bm{r}\right), \quad\bm{g}_{n}=\delta\cdot\bm{G}_{n}-\theta \hat{z}\times\bm{G}_{n};\nonumber  \\
    \varphi^h_{\rm W} = U_0\sum_{n=1,2,3}\left[\sin\left(\bm{g}_n\bm{r}\right) - \frac{1}{2}\cos\left(\bm{g}_n\bm{r}\right)\right], \qquad\qquad\quad
\end{align}
with $U_0\approx 12$ meV for diselenides and $U_0\approx 15$ meV for disulfides. 

We note that the above-mentioned lateral piezocharge transfer does not affect the energy of the bound electron-hole pairs such as  interlayer excitons ({\it iX}) \cite{SzyniszewskiPRB2017}, whereas the indirect band gap modulation by vertical (MoX$_2$ $\to$ WX$_2$) charge transfer does.  In twistronic AP-bilayers, this sets the {\it iX} energy minimum in the XX corners of the moir\'e supercell, see Fig. \ref{Fig4}C, typically, $\sim20$\,meV lower than in 2H and WMo areas for WSe$_2$/MoSe$_2$, and $\sim30$\,meV for WS$_2$/MoS$_2$. At low temperatures, these XX corners of the domain wall network would act as traps for both bright and dark interlayer excitons, potentially  providing conditions for accumulation of dark {\it iX} (traced to the spin-valley splitting \cite{Kormnyos2015,WangPRL2015}) due to their lower energy, or even create conditions for the single-photon emission.   

\begin{figure}
	\includegraphics[width=\columnwidth]{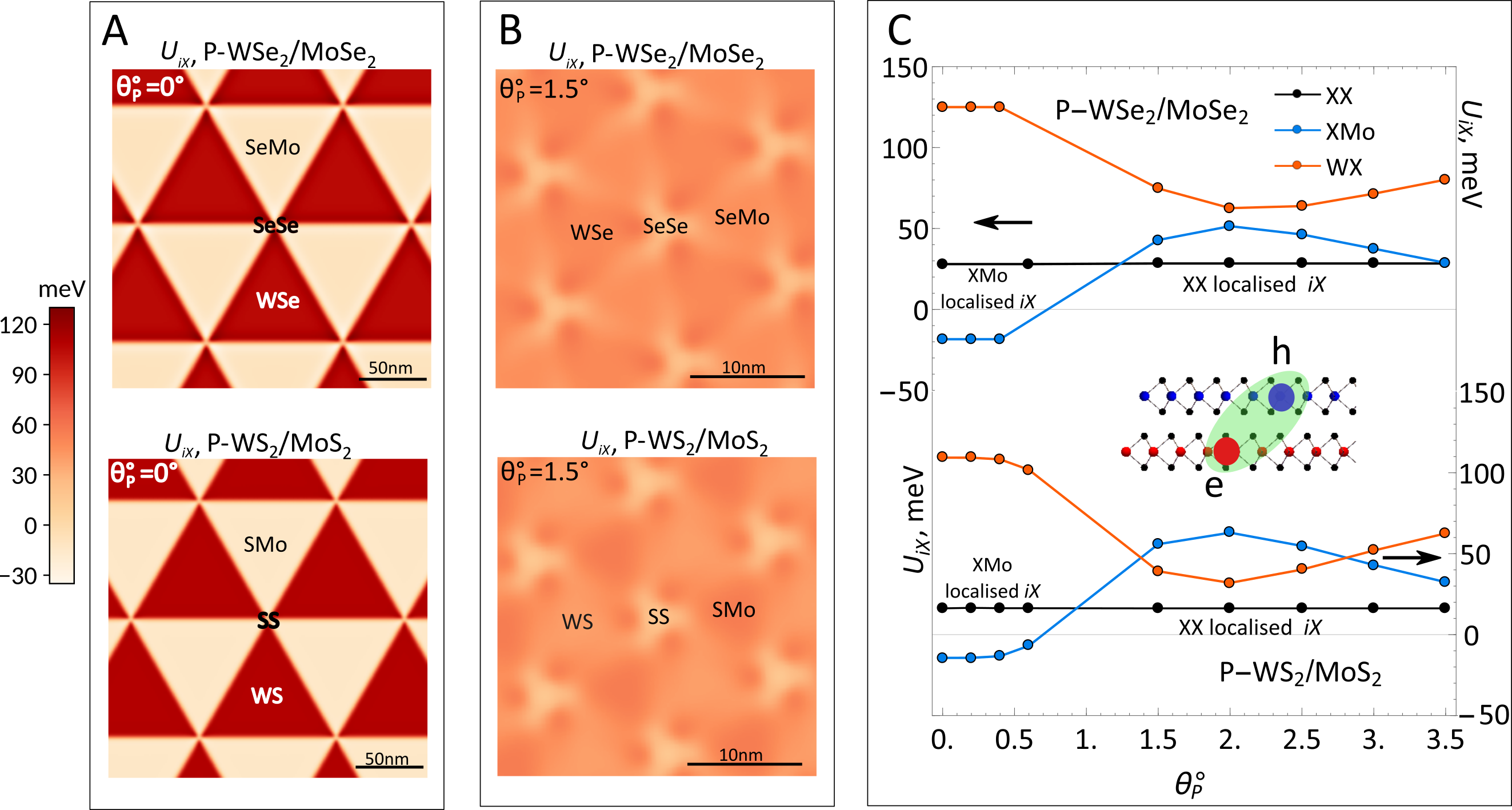}
	{\small\caption{\label{Fig5} {\bf Energy landscapes for interlayer excitons ({\it iX}) in  P-WSe$_2$/MoSe$_2$ (top) and P-WS$_2$/MoS$_2$ (bottom).}
		{\bf A.} For $\theta_P^{\circ}=0^{\circ}$ {\it iX} are confined inside triangular XMo domains. {\bf B.} For $\theta_P^{\circ}=1.5^{\circ}$ minima of {\it iX} energy appear in XX regions. On maps {\bf A} and {\bf B} horizontal and vertical directions correspond to zigzag and armchair crystallographic directions of constituent monolayers, respectively. {\bf C.} Twist-angle dependence of {\it iX} energy in XX, XMo, and WX regions of moir\'e supercell, with {\it iX} energy minima shifting from XMo domains to XX vortices at $|\theta_P^{\circ}| \approx 1^{\circ}$.	
	}}
\end{figure}

In contrast to the AP orientation, in P-bilayers the contributions of piezoelectric and ferroelectric charge transfers produce similar effects. This is because the piezocoefficients in monolayers with the same orientation of unit cells have the same sign, $e_{11}^{\rm W}\approx e_{11}^{\rm Mo}$, and as $\bm{u}^{\rm W}\approx -\bm{u}^{\rm Mo}$, leading to opposite signs of the piezocharges in top/bottom layers \cite{Enaldiev_PRL} and piezopotentials $\phi_{\rm Mo}\approx-\phi_{\rm W}$. Consequently, the combined effect of piezo- and ferroelectric contributions ({\it i.e.} charge double layer) electrostatically modulates the layer-indirect band gap in the type-II heterostructure WX$_2$/MoX$_2$. and the {\it iX} energy,  $U_{{\it iX}}$. To mention, for the AP-heterostructures the lowest spin-valley states of {\it iX} are bright, according to the optical transition selection rules \cite{Kormnyos2015,WangPRL2015,DanovichPRB2018}. The resulting energy landscape for the bright ground state of {\it iX} is shown in Fig. \ref{Fig5}A, indicating that the XMo regions form boxes (with almost the same depth $\approx 130$\,meV for sulphides and selenides) accumulating {\it iXs} in heterostructures with $\theta_{P}^{\circ} \lesssim 1^{\circ}$. As for P-bilayers we neglected additions to the {\it iX} energy from the off-resonant interlayer hybridisation of (conduction-conduction and valence-valence) band edge states (see Methods). The localisation of layer-indirect excitons in XMo domains is determined by the dominant role of the ferroelectric interlayer polarisation in such marginally twisted structures, whereas for twist angles $\theta_P^{\circ} \gtrsim 1.2^{\circ}$, the piezoelectric polarisation of domain walls between 3R stacking domains pushes the {\it iX} energy minima towards XX-stacking sites of triangular domain wall network, Fig. \ref{Fig5}B. For bilayers with even larger angles ($\theta_{P}^{\circ}\gtrsim 3^{\circ}$), where the lattice reconstruction is weak, the {\it iX} energy difference between XMo and XX areas progressively disappears, Fig. \ref{Fig5}C, opening channels for {\it iX} propagation across the supercell. In the latter regime we expect the formation of dispersive minibands for {\it iX}s, as discussed in Ref. \cite{ruiz2019interlayer}. Similarly, the crossover of {\it iX} confinement from XMo domains to XX stacking areas at $1^{\circ}<|\theta_P^{\circ}|\lesssim 1.2^{\circ}$ may also take a form of delocalised dispersive excitons.  

Overall the proposed theory establishes a rich variety of confinement conditions for electrons, holes and layer-indirect excitons across moir\'e superlattice in twistronic type-II heterostructures WX$_2$/MoX$_2$. The richness of predicted behaviour stems from the interplay between piezoelectric and weak ferroelectric effects. In contrast to homobilayers \cite{Edelberg2020}, the interplay between lattice reconstruction and charge transfer in slightly twisted same-chalcogen heterobilayers makes localisation of individual charge carriers and optically active excitations in AP-bilayers sensitive to a small variation of the twist angle, in the range of $\theta_{AP}^{\circ}\sim 0.2^{\circ}$. While such a sensitivity may represent a long-term opportunity for tuning the electronic transport properties and photoemission dynamics of such systems, in the first instance, it suggests that a homogeneity of the twist angle with accuracy better than $0.1^{\circ}$ will be necessary for achieving a systematic reproducibility in the  optoelectronic characteristics of WX$_2$/MoX$_2$ heterostructures.       

\section*{Methods}

{\bf Multiscale modeling}  \cite{jung2015origin,Enaldiev_PRL} in the reported study exploits the fact that the moir\'e superlattice period  $\ell\approx a/\sqrt{\theta_{P,AP}^2+\delta^2}$ is orders of magnitude longer (for $|\theta_{P,AP}^{\circ}|\leq 6^{\circ}$ which corresponds to $|\theta_{P,AP}|< 0.1$) than the lattice constant, $a$, of each monolayer. Because of such length scale separation local properties of each small area of a moir\'e supercell are quantified by performing {\it ab initio} DFT computations for ideally aligned pair of monolayers, however, with the locally defined lateral offset $\bm{r}_0$ and interlayer distance, $d$, which vary across the supercell. For convenience we set $\bm{r}_0=0$ for XX stacking (coinciding lateral positions of all chalcogens) for both P and AP orientations of the crystals. For the DFT computations we considered commensurate monolayers and used van der Waals corrected code (optB-88vdW) neglecting spin-orbit coupling and setting a plane-wave cutoff of 816.34 eV (60 Ry). The separation of micro- and mesoscopic scales is accounted for by $\bm{r}_0(\bm{r})=\delta \bm{r}+\theta_{P,AP}\hat{z}\times\bm{r} + \bm{u}^{\rm W}-\bm{u}^{\rm Mo}$, where the first term originate from a small difference between MoX$_2$ and WX$_2$ lattice constants, $\delta=2\left(a_{\rm Mo}-a_{\rm W}\right)/\left(a_{\rm Mo}+a_{\rm W}\right)$, the second term takes into account a fixed twist angle, and the last comes from lateral deformations in each layer bringing their lattices to the globally energetically favourable state. Displacements $\bm{u}^{\rm W/Mo}$ are related to strain in each layer $2u_{ij}^{\rm W/Mo}=\partial_iu^{\rm W/Mo}_j+\partial_ju^{\rm W/Mo}_i$, and  local interlayer distance, $d$, varies across the supercell adjusting to the optimal value for each local stacking configuration. 

We find the energetically favourable global state of the superlattice by minimising the total energy of the deformed crystal which includes the interlayer adhesion energy, $W_{P/AP}(\bm{r}_0,d)$, and the energy cost of lateral deformations. To cross from the microscopic scales (at which the adhesion energy is computed using DFT) to the mesoscale, we implement an interpolative analytical description for $W_{P/AP}(\bm{r}_0,d)$ and use it to formulate an effective energy density functional \cite{Enaldiev_PRL}. Numerical minimisation of $\mathcal{E}$ on a fine grid produces a detailed structure of the strain field and interatomic distances across the moir\'e supercell. This information is, then, used to compute piezoelectric charges, generated by inhomogeneity of strain and the interlayer charge transfer due to a weak ferroelectric effect. This last step requires additional DFT calculations, which are performed for the same set of stacking configurations, as in the analysis of adhesion energy and used to build an analytical interpolation formula for the potential step  $\Delta(\bm{r}_0,d)$ caused by the interlayer charge transfer. Finally, the computed $\bm{r}_0(\bm{r})$, $d(\bm{r}_0(\bm{r}))$, on-layer piezopotentials $\phi_{\rm W}$ in WX$_2$ and $\phi_{\rm Mo}$ in MoX$_2$, and $\Delta$ are combined together to compute maps displayed in Figs. \ref{Fig0},\ref{Fig4}, and \ref{Fig5}.  

{\bf Interpolation formula for adhesion energy $W_{P/AP}$.} First of all, we implement DFT to compute adhesion energies of two aligned and commensurate WX$_2$ and MoX$_2$ monolayers for several lateral offsets corresponding to stacking arrangements, shown in the insets in Fig. \ref{Fig2}, and various interlayer distances. The commensurability of the monolayer lattices is attained by in-plane stretching/shrinking of one of the monolayers to the lattice of the other, keeping X-X intersublayer distance in each layer fixed to its monolayer value. Due to a small lattice mismatch between WX$_2$ and MoX$_2$ ($\delta=0.3\%$ for X=Se and $0.2\%$ for X=S), the DFT results for adhesion energy (Fig. \ref{Fig2}A,B) are independent of which way such commensurability is imposed \cite{Enaldiev_PRL}. An analytical description of the interlayer adhesion energy as a function of stacking is achieved by using the Fourier expansion of $W_{P/AP}(\bm{r}_0,d)$ for a fixed $d$, keeping track of the constant term and six harmonics from the first star of reciprocal lattice vectors, $\pm \bm{G}_{1,2,3}$ ($|\bm{G}_{n}|=G$) of the monolayer. The resulting interpolational formula reads, 
\begin{equation}\label{Eq:adhesion}
\begin{split}
     W_{P/AP}(\bm{r}_0,d) \approx f(d_0) + \varepsilon z^2+
    w_1\left[1-Q z\right]\sum_{n=1,2,3}\cos\left(\bm{G}_n\bm{r}_0\right) \nonumber\\ +w_2\left[1-G z\right]\sum_{n=1,2,3}\sin\left(\bm{G}_n\bm{r}_0+\gamma_{P/AP}\right). \nonumber
\end{split}
\end{equation}
Here, $z=d-d_0$ is a deviation of the interlayer distance from the minimum of $f$ (configuration-averaged $W(d)$ coinciding with $W(d)$ for the 6$^{\rm th}$ configuration in Fig. \ref{Fig2}A,B), so that $f\approx f(d_0)+\varepsilon z^2$ is an expansion of $f$ around its minimum, $d_0$; $\gamma_{AP}=0$ and $\gamma_{P}=\pi/2$, and $\varepsilon$, $w_{1,2}$, $Q$ are fitting parameters listed in Table \ref{tab_Fit}. This offers a simplified version of the interpolation formula used in Ref. \cite{Enaldiev_PRL}. To account for out-of-plane relaxation of WX$_2$ and MoX$_2$ we minimise $W_{P/AP}$ with respect to $z$, which enables us to find a local relation $z=Z(\bm{r}_0)$. Note that the interlayer distance variation, $Z(\bm{r}_0)$, from the lowest to the highest energy stackings, $\sim d_0\pm 0.3$\AA, is much smaller than interlayer distance itself, which justifies the expansion implemented in $W_{P/AP}$. Also, as it has been shown earlier \cite{Enaldiev_PRL}, the energy cost of bending deformations of WX$_2$ and MoX$_2$ monolayers is negligibly small (compared with that of lateral strain and variations of adhesion energy) due to the long range of the deformations of these atomically thin membranes (the same approximation was used in Ref.\cite{jung2015origin} when analysing graphene-hBN heterostructures).  Overall this determines the locally adjusted energy, dependent only on the local interlayer offset $\bm{r}_0$ as
\begin{align}\label{Eq:adhesion_2}
    \widetilde{W}_{P/AP}(\bm{r}_0) = -\varepsilon Z^2(\bm{r}_0) + 
    \sum_{n=1}^{3}\left[w_1\cos\left(\bm{G}_n\bm{r}_0\right) + w_2\sin\left(\bm{G}_n\bm{r}_0+\gamma_{P/AP}\right)\right], \nonumber \\
    Z(\bm{r}_0)=\frac{1}{2\varepsilon}\sum_{n=1}^3\left[ w_1Q\cos\left(\bm{G}_n\bm{r}_0\right) + w_2G\sin\left(\bm{G}_n\bm{r}_0+\gamma_{P/AP}\right)\right]. \nonumber
\end{align}

{ \bf Mesoscale lattice reconstruction in twisted bilayers.}  Having defined a locally adjusted adhesion energy function, $\widetilde{W}_{P/AP}$, we minimise the total energy,
\begin{equation}\label{Eq:min_functional}
\mathcal{E}=\int_{\rm supercell} d^{2} \boldsymbol{r}\left\{\sum_{l={\rm W,Mo}}\left[\frac{\lambda_l}{2}\left(u_{ii}^{(l)}\right)^{2} + \mu_l u_{ij}^{(l)}u_{ji}^{(l)}\right] + \widetilde{W}_{AP/P}\left(\bm{r}_0(\bm{r})\right)\right\},\nonumber
\end{equation}
with respect to displacement fields $\bm{u}^{\rm W}$ and $\bm{u}^{\rm Mo}$ implicit in $\bm{r}_0(\bm{r})=\delta \bm{r}+\theta\hat{z}\times\bm{r} + \bm{u}^{\rm W}-\bm{u}^{\rm Mo}$. Here  $\lambda$, $\mu$ are elastic moduli, which values are listed in Table \ref{tab_Fit}. To compute the optimal distributions of the displacement fields $\bm{u}^{\rm W}(\bm{r})$ and $\bm{u}^{\rm Mo}(\bm{r})$, which we keep periodic, with the moir\'e superlattice period, we solve a system of Lagrange-Euler equations using the finite difference method. Here we use a homogeneous grid aross a moir\'e supercell to solve a system of non-linear algebraic equations, supplemented by periodic boundary conditions based on the interior point method, as implemented in the GEKKO Optimization Suite \cite{Beal_2018}. 

{\bf DFT analysis of interlayer charge transfer.} The analysis of weak ferroelectricity in this work employs the \textsc{VASP}
package \cite{VASP} with projector augmented
wave (PAW) pseudopotentials. We approximated the exchange correlation functional using the generalised gradient approximation (GGA) of Perdew, Burke and Ernzerhof \cite{PBE}. 
The cutoff energy for the plane-waves was set to 600~eV and the in-plane Brillouin zone sampled by a
$12 \times 12$ grid. The in-plane lattice constants of the constituent monolayers are strained to the mean of their experimental values, taken from Refs. \cite{MoX2_experimental_structures} and \cite{WX2_experimental_structures} for MoX$_2$ and WX$_2$, respectively, while the intralayer chalcogen-chalcogen distances are left unchanged for each constituent monolayer. In the inset in Fig. \ref{Fig2}D we illustrate the DFT-computed behaviour of the difference between the plane-averaged electron potential energy for a WSe$_2$/MoSe$_2$ heterobilayer (for 3R stacking and interlayer distance $d=6.477$\AA), and that from the sum of the potential energies from isolated individual WSe$_2$ and MoSe$_2$ monolayers. To avoid having to artificially resolve the mismatch in potential energy which would arise at the boundary of a supercell containing a single heterobilayer, results for heterobilayers in this work were computed using a supercell containing two mirror-reflected images of the structure.

It should be noted that the van der Waals corrected DFT (optB-88vdW) used in determining the interlayer adhesion energies \cite{Enaldiev_PRL} gives a slightly larger optimal interlayer distance ($6.56$\AA\ for WSe$_2$/MoSe$_2$, and $6.25$\AA\ for WS$_2$/MoS$_2$) for 2H/3R stacking configuration in WX$_2$/MoX$_2$ bilayers than the values found in experiment for the corresponding homobilayers ($6.46$\AA, $6.48$\AA, $6.15$\AA\, and $6.17$\AA\, for MoSe$_2$, WSe$_2$, MoS$_2$, and WS$_2$, respectively \cite{MoX2_experimental_structures,WX2_experimental_structures}). However, usage of a linear fit for $\Delta_0$ and $\Delta_a$ with relative interlayer distance dependence $d-d_0$ allows us to take the difference into account by counting the variations of $\Delta$ from a shifted interlayer distance $d-d_0\to Z(\bm{r}_0(\bm{r}))-0.1$\AA.

{\bf Piezo- and ferroelectric potentials for electrons and holes induced by piezo- and ferroelectric charge transfers.} Here we combine the potential jump $\Delta$ at the double layer induced by the weak ferroelectric effect with the potential created by piezoelectric charges, accumulated around domain walls in the reconstructed moir\'e superlattice. In the analysis of the latter we also take into account that the electric field generated by combined piezocharges of AP-oriented WX$_2$ and MoX$_2$ layers is screened by the dielectric environment, in, e.g., hBN-encapsulated bilayers. In particular we take into account the in-/out-of-plane anisotropy of hBN dielectric parameters (\cite{laturia2018dielectric} \mbox{$\epsilon_{||}=6.93$} and \mbox{$\epsilon_{\perp}=3.76$}). Additionally, for both P and AP orientations we take into account the in-plane dielectric polarisability of WX$_2$ and MoX$_2$, $\alpha_{2D}^{\rm W/Mo}=d_0(\epsilon^{\rm WX/MoX}_{||}-1)/4\pi$ \cite{2dPolarizationPRB}, leading to the induced charges 
\begin{equation}
\rho^{\rm W/Mo}_{ind}=\alpha^{\rm W/Mo}_{2D}\bm{\nabla}^2\phi_{\rm W/Mo}, \nonumber 
\end{equation}
where $\epsilon^{\rm WX}_{||}$ and $\epsilon^{\rm MoX}_{||}$ are in-plane dielectric permittivities of bulk WX$_2$ and MoX$_2$ crystals, $\bm{\nabla}=(\partial_x,\partial_y)$, and $\phi_{\rm W/Mo}$ is the electrostatic piezopotential in the W/Mo layer. We note that piezocharge densities $\rho^{\rm W/Mo}$ are symmetric with respect to the mirror plane in each layer. Therefore, the piezocharges in one monolayer are screened only by the out-of-plane polarised charges in the opposite layer. Under this assumption the magnitudes of out-of-plane polarised charges are 
\begin{equation}
\sigma_{\rm W/Mo}=\frac{\left(\epsilon_{\perp}^{\rm WX/MoX}-1\right)}{2\epsilon_{\perp}^{\rm WX/MoX}}\rho^{\rm Mo/W},    \nonumber
\end{equation}
where $\epsilon_{\perp}^{\rm WX}$ ($\epsilon_{\perp}^{\rm MoX}$) is the out-of-layer dielectric permittivity of bulk WX$_2$ (MoX$_2$) crystal. Here, we used the following values of dielectric permittivities $\epsilon_{||}^{\rm WX}=14.6/16$, $\epsilon_{||}^{\rm MoX}=16.3/17.9$ and $\epsilon_{\perp}^{\rm WX}=6.7/7.9$, $\epsilon_{\perp}^{\rm MoX}=7.3/8.5$ known for tungsten and molybdenum disulfide/diselenide bulk crystals \cite{laturia2018dielectric}.  In our electrostatic model planes hosting charges $\sigma_{\rm W/Mo}$ are distant from the mirror plane in corresponding layer by distance $d_0/2$, with density $+\sigma_{\rm W/Mo}$ from outer and $-\sigma_{\rm W/Mo}$ from inner sides of the bilayer structure (here we assume $d_0$ is interlayer distance). Finally, we find $\phi_{\rm W/Mo}$ by solving the Poisson equation with appropriate boundary conditions on the planes with the charges discussed above.  

The total potential energy of charge carriers in MoX$_2$ and WX$_2$ layers is a sum of the piezo- and ferroelectric contributions, where the ferroelectric part is determined by $\frac{1}{2}\left[\Delta(\bm{r}_0(\bm{r}))-\Delta_0(d_{2H/3R})\right]$.  Thus, for potential energy of electrons and holes shown in Figs. \ref{Fig4} we use $\varphi_{\rm Mo}^e=-e\phi_{\rm Mo}+\frac{1}{2}\left[\Delta-\Delta_0(d_{2H/3R})\right]$ and $\varphi_{\rm W}^h=e\phi_{\rm W} + \frac{1}{2}\left[\Delta-\Delta_0(d_{2H/3R})\right]$, respectively. Then {\it iX} energy (Fig. \ref{Fig5}) is their sum $U_{iX}=-e(\phi_{\rm Mo}-\phi_{\rm W})+\Delta$. 

Finally, we note that in calculation of energy profiles for electrons/holes (Fig. \ref{Fig4}) and {\it iXs} (Fig. \ref{Fig5}) we neglect the energy modulation resulting from interlayer hybridisation of conduction-conduction and valence-valence band edge states. This is because in type-II heterostructures WX$_2$/MoX$_2$ the band edges are non-resonant, with \cite{APL2013,band_alignPRB2016,kang2013band} $\sim 500$\,meV offsets in conduction and valence bands, which leads only to small corrections in the second order of perturbation theory. Therefore, its influence on the band edges in WX$_2$/MoX$_2$ bilayers should be much smaller than in TMD homobilyers where states are resonantly coupled, and where characteristic tunneling matrix element is, anyway, smaller ($\sim 20$\,meV \cite{YaoPRB2017,ruiz2019interlayer,alexeev2019resonantly}) than the effects discussed above. 

{\bf Aknowledgements.}  We thank Mingxin Chen, Andre Geim, Roman Gorbachev, Hongkun Park, Alex Summerfield, David Ruiz-Tijerina, Wang Yao, and Hongyi Yu for useful discussions. This work has been supported by EPSRC grants EP/S019367/1, EP/S030719/1, EP/N010345/1, EP/V007033/1; ERC Synergy Grant Hetero2D; Lloyd’s Register Foundation Nanotechnology grant; European Graphene Flagship Project, and EU Quantum Technology Flagship project 2D-SIPC. V.V.E. (piezo- and ferrocharge model) acknowledges the support of the Russian Science Foundation (project no. 16-12-10411). 

\bibliographystyle{nature}
\bibliography{references}

\end{document}